\documentclass[pre,showpacs,aps,preprint]{revtex4}
\input epsf

\usepackage{times}
\usepackage{amstext,amsmath,amsfonts}

\begin{document}

\title{Gradient Networks}

\author{
Zolt\'an Toroczkai$^{1}$,
Bal\'azs Kozma$^{2}$,
Kevin E. Bassler$^{3}$,
N. W. Hengartner$^{4}$ and
G. Korniss$^{2}$}
\affiliation{
$^{1}$Center for Nonlinear Studies, Theoretical Division, Los
Alamos National Laboratory, MS B258 Los Alamos, NM 87545, USA,  {\tt
toro@lanl.gov}, \\
$^{2}$Department of Physics, Applied Physics, and
Astronomy, Rensselaer  Polytechnic Institute, 110, 8$^{th}$ Street, Troy,
NY 12180, USA\\
$^{3}$Department of Physics, 617 Science and Research
bld. I, University of Houston, Houston, TX, 77204, USA\\
$^4$ Statistics Group, Decisions Applications Division, Los
Alamos National Laboratory, MS F600 Los Alamos, NM 87545, USA
}

\date{\today}

\begin{abstract}
We define gradient networks as directed graphs formed by local gradients of
a scalar field distributed on the nodes of a substrate network $G$.
We derive an exact
expression for the in-degree distribution of the gradient network when the
substrate is a binomial (Erd\H{o}s-R\'enyi) random graph, $G_{N,p}$.
Using this expression we show that
the in-degree distribution $R(l)$ of gradient graphs on $G_{N,p}$ obeys the
power law $R(l)\propto l^{-1}$ for arbitrary, i.i.d. random scalar fields.
We then relate gradient graphs to congestion tendency in network flows
and show that while  random graphs become
maximally congested in the large network size limit, scale-free networks are  not,
forming fairly efficient substrates for transport.
Combining this with other constraints, such as uniform edge cost, we obtain
a plausible argument in form of a selection principle, for why a number of
spontaneously evolved massive networks are scale-free.
This paper also presents detailed derivations
of the results reported in Ref. \cite{TB04}.
\end{abstract}


\maketitle

\section{Introduction}

It has recently been recognized \cite{AB02,N03,DM03} that a large number of
systems are organized into structures best described by complex networks,
or massive graphs. Many of these networks, also called {\em scale-free networks},
such as citation networks \cite{R98},
the www \cite{AJB00}, the internet \cite{FFF99},
cellular metabolic networks \cite{JTAOB00,W01}, the sex-web \cite{LESAA03},
the world-wide airport network \cite{GMTA03,GA04}, and
alliance networks in the U.S. biotechnology industry \cite{PWKO03},
possess power-law degree distribution, $P(k) \sim k^{-\gamma}$
\cite{BA99}.
Scale-free networks are very different from pure random graphs, which
are well studied in the mathematical literature \cite{B01},
and which have ``bell curve'' Poisson degree distributions.
Therefore, it is natural to ask: Why do scale-free networks
emerge in nature?

The diverse range of systems
for which scale-free networks are important
suggests that perhaps
there is a simple common reason for their development.
Generally, real-world networks do not form or evolve simply by purely random
processes. Instead, networks develop in order to fulfill a {\em main function}.
Often that function is to {\em transport} entities such as information, cars,
power, water, forces, etc.
It is thus plausible that the structure that the network develops
(scale-free in particular) will be one that ensures efficient transport.
Recent studies that explore the connection between network topology and
flow efficiency were done by a number of researchers, including Valverde and Sol\'{e}
\cite{VS04} in the context of the internet, by Valverde Ferrer Cancho and Sol\'{e}
\cite{VFS02} in the context of software architecture graphs,
and Guimer\`{a} et.al. \cite{GMTA03} in the context of the world-wide airport network.
All these studies clearly show that the flow optimization dynamics which attempts
to maintain the overall efficiency, will induce strong constraints on
the structural evolution of the network.

In this paper, we investigate the processing efficiency of flows on
networks when the flows are generated by gradients of a
scalar field distributed on the nodes of a network.
This approach is motivated by the idea that transport processes are often
driven by local gradients of a scalar. Examples include
electric current which is driven by a gradient of electric potential,
and heat flow which is driven by a gradient of temperature.
The existence of gradients has been also shown to play an important role
in biological transport processes, such as cell migration
\cite{N84}: chemotaxis,  haptotaxis, and galvanotaxis (the later
was shown to play a crucial role in morphogenesis).

Naturally, the same mechanism will generate flows in complex networks as well.
Besides the obvious examples of traffic flows, power distribution on the
grid, and waterways, we present two less known examples of systems where
gradient-induced transport
on complex networks plays an important role:
1) Diffusive load balancing schemes used in
distributed computation \cite{RSW98} (and also employed in
packet routing on the internet), and  2)
Reinforcement learning on social networks with competitive dynamics
\cite{ATBK03}.
In the first example, a computer (or a router) asks its neighbors
on the network for their current job load (or packet load), and then
the router balances its load with the neighbor that has the {\em minimum} number
of jobs to run (or packets to route).
In this case the scalar at each node is the negative of the number of
jobs at that node, and the {\em flow} occurs in the direction of
the gradient of this scalar in the node's network neighborhood.
In the second example, a number of agents/players who are all part
of a social network, compete in an iterated game
based with limited information \cite{ATBK03}.
At every step of the game each agent has to decide who's advice to follow
before taking an action (such as placing a bet), in its circle of acquaintances
(network neighborhood). Typically, an agent will try to follow
that neighbor which in the past proved to be the most reliable.
That neighbor is recognized using a reinforcement learning mechanism:
a score is kept for every agent measuring its past success at predicting the
correct outcome of the game, and then
each agent follows the advice of the agent in its network neighborhood
which has the {\em highest score} \cite{ATBK03} accumulated up to that point in time.
In this case, the scalar is the past success score kept for each agent.

The remainder of this paper is organized as follows.
In Section II
we systematically build a framework
for analyzing the properties
of gradient flows on networks, which, as it will be demonstrated,
generically organizes itself into a
directed network structure without loops.
In Section III, we obtain the exact expression
for the in-degree distribution of the gradient flow network on
binomial random graphs
and show that in a certain scaling limit the gradient flow network becomes
a scale-free network.
We also discuss possible connections of this result to
sampling biases in trace-route measurements
that have been used to infer the topology of the internet.
Finally, in Section IV we study how the structure of a network
affects the efficiency of its transport properties,
and offer a possible explanation in the form of a selection principle for
the emergence of real-world scale-free networks.

\section{Definition of a Gradient Network}

Let us consider that transport takes place on a fixed network $G=
G(V,E)$ which we will call in the remainder, the {\em substrate graph}.
It has $N$ nodes, $V=\{0,1,...,N-1\}$ and the set of edges $E$ is specified
by the adjacency matrix ${\bf A} = \{a_{ij}\}$ ($a_{ij} = 1$ if $i$ and $j$
are connected, $a_{ij} = 0$ otherwise, and $a_{ii}=0$). Given a node
$i$, we will denote its set of neighbors in $G$ by $S_{i}^{(1)}
=\{j\in V \;|\; a_{ij} = 1\}$.
Let us also consider a
scalar field (which could just as well be called `potential landscape')
${\bf h} = \{h_0,..,h_{N-1}\}$ defined on the set of nodes $V$,
so that every node $i$ has a scalar value $h_i$ associated to it.

\begin{figure}[htbp]
\protect\vspace*{-0.1cm} \epsfxsize = 3.0 in
\centerline{\epsfbox{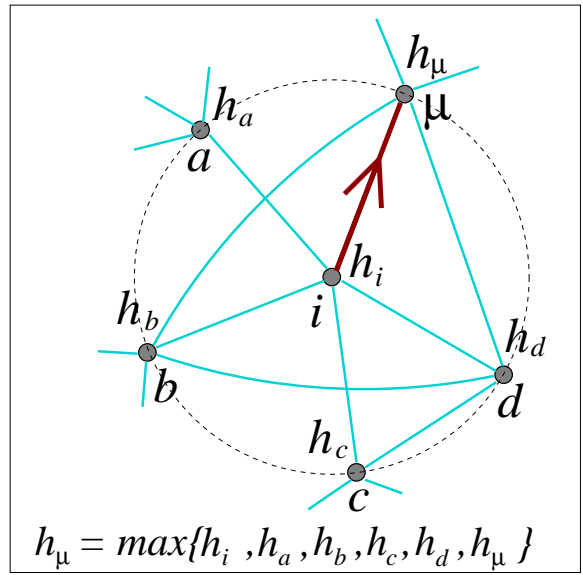}}
\protect\vspace*{-0.2cm}
\caption{Definition of a  gradient on a network.
The gradient at node $i$ is a directed edge pointing towards
the maximum value of the scalar (node $\mu$) in the node's neighborhood.}
\label{fig:BP_Fig1}
\end{figure}

We define the {\em gradient} $\nabla h_i$ of the field {\bf h} in the node $i$
to be the {\em directed edge} $\nabla h_i = (i,\mu(i))$
which points from $i$ to that neighbor, $\mu(i) \in S_{i}^{(1)}\cup\{i\}$
on $G$ at which the scalar field has the maximum value in
$S_{i}^{(1)}\cup\{i\}$, i.e.:
\begin{equation}
\mu(i) = \underset{j \in S^{(1)}_i
\cup \{ i \}}{\mbox{argmax}}(h_j), \label{argmax}
\end{equation}
see Fig. \ref{fig:BP_Fig1}.
According to its classical definition, a gradient vector points in the
direction of the steepest ascent at a point on a continuous ($d$-dimensional)
landscape. The above definition  is a natural generalization to the case
when the continuous landscape is replaced by a graph.

Note that $\mu(i) = i$, if $i$ has the largest scalar value in its neighborhood
(i.e. in the set $S_{i}^{(1)}\cup\{i\}$), and in this case the gradient edge
is a {\em self-loop} at that node. Since
${\bf h}$ always has a global maximum, there is always at least one self-loop.
It is possible that Eq. (\ref{argmax}) has more than one solution (several equal
maxima) in the case of which we say that the scalar field is degenerate.
In this paper we deal only with non-degenerate fields, which is typical when
for example ${\bf h}$ is a continuous stochastic variable.

This allows us to define: the set ${\bf F}$ of
gradient edges on $G$, together with the vertex set $V$ form the {\em Gradient
Network}, $\nabla G = \nabla G(V,{\bf F})$.

Assuming that all
edges have the same `conductance', or transport properties,
the gradient network will be the substructure of the
original network which at a given instant will {\em channel the bulk of the flow},
and thus alternatively can be called as the {\em maximum flow subgraph}.

In general, the scalar field will be evolving in time, due to the gradients
generated currents, and also to possible external sources and sinks on the network
(for example packets are generated and used up at nodes, but they can also be lost).
As a result, the gradient network $\nabla G$ will be {\em time-dependent,
highly dynamic}.

\subsection{Some general properties of Gradient Networks}

\begin{figure}[htbp]
\protect\vspace*{-0.1cm} \epsfxsize = 5.0 in
\centerline{\epsfbox{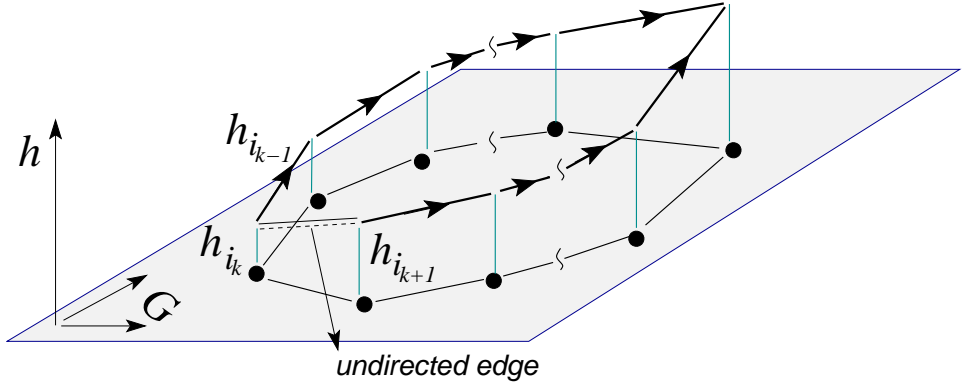}}
\protect\vspace*{-0.2cm}
\caption{There cannot be loops in a non-degenerate $\nabla G$.}
\label{fig:gnFig2}
\end{figure}

Here we will first show a number of fundamental
{\em structural} properties valid for all {\em instantaneous} gradient networks,
and then study the degree distribution of gradient networks generated by stochastic
scalar fields on random graphs,
and scale-free networks. This will lead us to show
that scale-free networks are more efficient substrates for transport than random
graphs.
The first important observation we make about gradient networks is:\\
\noindent{\em Non-degenerate gradient networks form forests
(i.e., there are no loops in $\nabla G$, and  it is a union of trees, more
exactly of in-directed, planted pines).}\\
To prove this statement, assume that on the contrary, there is a closed path
$\gamma = \{\nabla h_{i_1}, \nabla h_{i_2},..., \nabla
h_{i_m} \}$, $m \geq 3$ made up only of
directed edges from $\bf F$, see Fig. \ref{fig:gnFig2}.
Let $i_k$ be the node on this path for which $h_{i_k} = \min \{
h_{i_1},h_{i_2},...,h_{i_m}\}$. Node $i_k$ has exactly two
neighbors on $\gamma$, nodes $i_{k \pm 1}$, but only one gradient
direction, $\nabla h_{i_k}$ pointing away from $i_k$. Since
$h_{i_{k\pm 1}} > h_{i_k}$, none of the neighbors $i_{k \pm 1}$
will have their gradient edges pointing into $i_{k}$. Since there are
two edges, $(i_k,i_{k-1})$ and $(i_{k},i_{k+1})$, and only one
gradient edge from $i_k$, one of the edges must not be a gradient
edge, and thus the loop is not closed, in contradiction with the
assumption that $\gamma$ is a loop with only gradient edges.
Using a similar reasoning we can show that for non-degenerate scalar fields,
there is no continuous path in $\nabla G$ connecting two local
maxima of the scalar field ${\bf h}$. This means that on a given
tree of $\nabla G$ there is only one local maximum of the scalar,
and it is the only node with a self-loop
on that tree. As a consequence, {\em the number of
trees in the forest equals the number of local maxima
of the scalar field ${\bf h}$ on $G$}.  The fact that $\nabla G$ is made
of trees (no loops), is rather advantageous for existing analytical
techniques, especially if we take into consideration that $\nabla G$
is the most important substructure driving the flow in the network.
Note that unless there is exactly one local maximum
(and thus global as well) of ${\bf h}$
on $G$, $\nabla G$ is disconnected into a number of trees and thus
$\nabla G$ is {\em not} a spanning tree.
Since every node has exactly one gradient direction from it, the out-degree
of every node on the gradient network is unity. It also means that $\nabla G$
has exactly $N$ nodes and $N$ edges (with at least one edge being a self-loop).
However, the {\em in-degree} of a node $i$, which is the number of gradient edges
{\em pointing into} $i$, can be anything in the range
$k^{(in)}_i \in \{0,1,..,k_i \}$, where $k_i$ is the degree of node $i$ on $G$.

\section{The in-degree distribution of a gradient network on
random graphs and random fields}

\begin{figure}[htbp]
\protect\vspace*{-0.1cm} \epsfxsize = 5 in
\centerline{\epsfbox{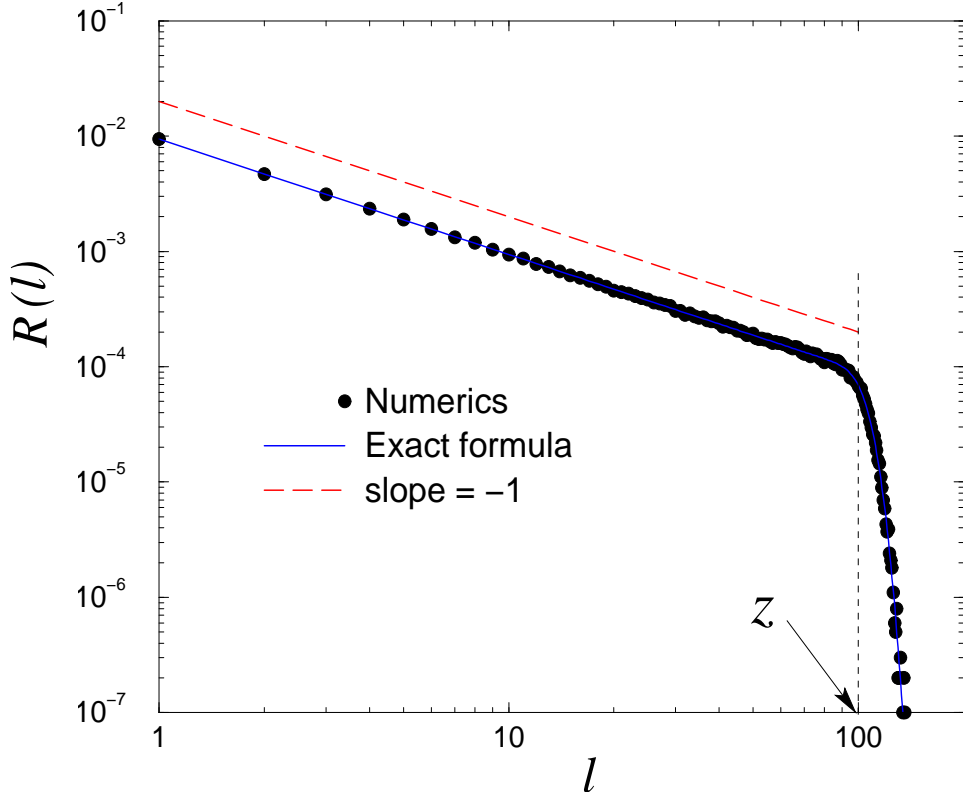}}
\protect\vspace*{-0.2cm}
\caption{Comparison between the exact formula (\ref{exact}) and numerics.
Here $N=1000$, $p=0.1$, ($z=100$). The numerical values are obtained
after averaging over $10^4$ sample runs.}
\label{fig:FIG4}
\end{figure}

In this section we show that when the substrate graph $G$
is a binomial random graph $G=G_{N,p}$ \cite{B01},
and ${\mathbf h}$ is an i.i.d. (meaning independent identically distributed)
random field over $V$, given by a distribution $\eta(h)$, the in-degree
distribution $R(l)=\mbox{Prob.}\{k^{(in)}_i=l\}$ of $\nabla G$ obeys the exact
expression:
\begin{eqnarray}
R(l) = \frac{1}{N}\sum_{n=0}^{N-1}\;
{N-1-n \choose l} \left[1-p(1-p)^n\right]^{N-1-n-l}
\left[ p(1-p)^n\right]^l\;,  \label{exact}
\end{eqnarray}
an expression {\em independent} on the particular form of the distribution
for the scalars, $\eta(h)$. The binomial random graph (also coined
in the physics literature as the Erd\H{o}s-R\'enyi random graph) is constructed by
taking all pairs $(i,j)$ of $N$ nodes and connecting them with probability $p$,
independently from other links. Figure \ref{fig:FIG4} shows the agreement
between numerical simulations and the above exact expression.
We will also show, that the gradient network $\nabla G$ becomes a
{\em scale-free (or power law) network} with
respect to the in-degree distribution,
in the scaling limit $N\to \infty$ and $p \to 0$, such
that $N p = z =const. \gg 1$.
The in-degree distribution in this limit
is described by the law:
\begin{equation}
R(l) \simeq \frac{1}{zl}\;,\;\;\;
0< l \leq z\;. \label{main}
\end{equation}
a behavior which is also apparent from Fig. \ref{fig:FIG4}.
 This power-law is a rather surprising result, since
the substrate graph is a random graph which is {\em not} scale-free, its
degree distribution (in the same limit) being Poisson, with a well defined
average degree $z$ (setting the scale) and faster than exponential decaying tails \cite{B01}.

A similar finding was reported in \cite{LBCX03} by
Lakhina et.al. by repeating the trace-route
measurements employed to sample the structure of the internet, on binomial
random graphs. Lakhina et. al. find that the spanning trees obtained this
way have a degree distribution that obeys the $1/k$ law.
Later, in Ref. \cite{CM04},
Clauset and Moore have presented an analytical approach to derive the $1/k$ law.
This suggests that perhaps there could be a mapping between the trace-route sampling
generated graphs and gradient networks. Although it is not an exact mapping, a close connection
can indeed be made, and this will form the subject of a forthcoming publication.
The main warning sign coming from the trace-route observations is then the fact that trace-route sampling
might not be the best way to measure the structure of the internet, given that on
random binomial graphs it fails miserably to reproduce its degree distribution (instead of
a Poisson, it gives a $1/k$ law). However, all is not lost, for the following reason:
trace-route measurements of the internet suggested a power-law dependence for its degree
distribution, $1/k^\gamma$ with an exponent of $\gamma$ taking values between 2 and 3, which
is definitely not close to unity! This excludes the binomial random graph as a model for the
internet. One might then wonder for what kind of graphs will trace-route measurements suggest
a power-law dependence with an exponent $\gamma > 2$?
In Section III, we make the observation that if the
substrate graph is a scale-free network with degree
distribution given by $1/k^\gamma$ ($\gamma > 2$) then
the corresponding gradient network {\em will also be a
scale-free network with the same exponent}
$\gamma$. Using the above mentioned close mapping between
trace-route trees and gradient networks
(namely, trace-route trees can be interpreted as suitably
constructed gradient networks) this
suggests that at least the assumption that the internet is
a power-law graph with exponent
$\gamma > 2$ is consistent with trace-route measurements.
Certainly, the problem of sampling biases
generated by trace-routes could be elucidated by answering
the following question: are there
non-power-law substrate graphs which would still generate
scale-free trace-route trees with exponent $\gamma > 2$?
This is an open question wordy of further investigation.

\section{Derivation of the exact expression}

In this section we give a combinatorial derivation for formula (\ref{exact}).
A more analytic and standard approach can be found in the Appendix, which
was our original method, and it has inspired the combinatorial one presented below.

In order to calculate the in-degree distribution $R(l)$ this way,
we first distribute the scalars on the node set $V$, then find those
link configurations which contribute to $R(l)$ when building the random graph
$G_{N,p}$ over these nodes.

\begin{figure}[htbp]
\protect\vspace*{-0.1cm} \epsfxsize = 2.3 in
\centerline{\epsfbox{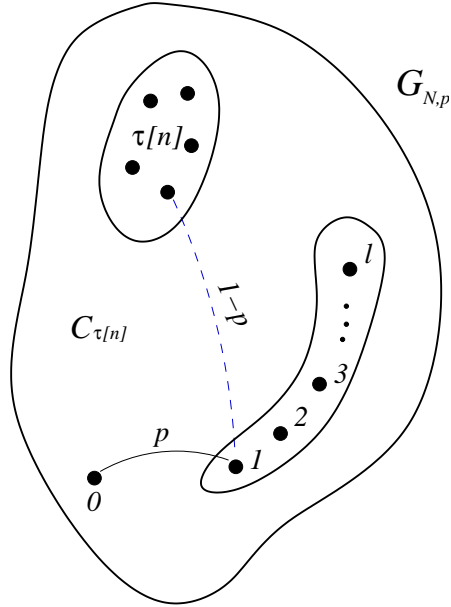}}
\protect\vspace*{-0.2cm}
\caption{Schematic of the construction given in the main text.}
\label{fig:Comb}
\end{figure}

Without restricting the generality we will calculate the distribution of in-links
for node 0. Let us consider a set of $n$ nodes from $V$, that does not contain node 0,
and it has the property that the scalar values at these nodes are all larger
than $h_0$.
We will denote this set by $\{\tau\}_n$. The complementary set of $\{\tau\}_n$ in
$V\setminus\{0\}$ will be denoted by $C_{\{\tau\}_n}$, see Fig. \ref{fig:Comb}.

In order to have {\em exactly} $l$ nodes pointing their gradient edges into node 0,
we must fulfill the following conditions: first, they have to be connected to node 0
and, second, they {\em must not} be connected to the set $\{\tau\}_n$ (otherwise, they
would be connected to a node with a scalar value larger than $h_0$, according to the
definition of $\{\tau\}_n$). The probability for one node to fulfill these two
conditions is $p(1-p)^n$, and since the links are drawn independently,
for $l$ nodes this probability is $\left[ p(1-p)^n\right]^l$. We must also require
that no other nodes will have their gradient links pointing into node 0. Obviously,
by definition, nodes from $\{\tau\}_n$ will not be pointing gradients into node 0.
Therefore, we have to make sure, that none of the remaining $N-1-l-n$ nodes from
$C_{\{\tau\}_n}$ will be pointing into 0. For one such node this will happen
with probability $1-p(1-p)^n$. For all the  $N-1-l-n$ such nodes this probability
will be $\left[ 1-p(1-p)^n \right]^{N-1-l-n}$. Thus, given a specific set $\{\tau\}_n$,
the probability of exactly $l$ in-links to node $0$ is:
\begin{equation}
{N-1-n \choose l} \left[ p(1-p)^n\right]^l
\left[ 1-p(1-p)^n \right]^{N-1-l-n}\;. \label{at}
\end{equation}
The combinatorial factor in (\ref{at}) counts the number of ways the set of $l$
nodes which point their gradient edges to node 0, can be chosen
from $C_{\{\tau\}_n}$.

The probability in (\ref{at}) was computed by fixing $h_0$ and
the set $\{\tau\}_n$. Next, we compute the probability $Q_n$ of
such an event for a given $n$, while letting the field
${\bf h}$ vary according to its distribution.
The probability for a node to have its scalar value larger than $h_0$ is:
\begin{equation}
\gamma(h_0)=\int\limits_{h_0} dh\; \eta(h)
\end{equation}
The probability to have exactly $n$ nodes with this property is given by:
\begin{equation}
 \left[\gamma(h_0) \right]^{n}
\left[ 1-\gamma(h_0)\right]^{N-1-n}\;.
\end{equation}
The number of ways the $n$ nodes can be chosen from $V\setminus\{0\}$ is
just the binomial ${N-1\choose n}$. Thus, the total probability $Q_n$ will
be given by:
\begin{equation}
Q_n = {N-1\choose n} \int dh_0\;\eta(h_0)\;\left[\gamma(h_0) \right]^{n}
\left[ 1-\gamma(h_0)\right]^{N-1-n}\;=\frac{1}{N}\;, \label{qun}
\end{equation}
where the last equality in the above is obtained after performing the change of variables
$du \equiv d\gamma(h_0)=dh_0\;\eta(h_0)$.

As a final step, by combining (\ref{qun})
with (\ref{at}), and summing over all possible $n$ values, we arrive at (\ref{exact}).

\subsection{Derivation of the $1/l$ scaling for the in-degree
distribution}

In order to obtain the scaling $1/l$ valid in the limit
$N \to \infty$, $p \to 0$, such that $z=pN=\mbox{const.}\gg 1$,
for $0 \leq l \leq z$,
we will use the saddle point method. We write equation (\ref{exact})
first in the form $R(l)=\frac{1}{N}\sum_{n=1}^{N}r_N(n,l)$ and then
exponentiate the argument. Using the Stirling's formula to the
first order ($\ln{(x!)}\approx x(\ln{x} - 1)$), one obtains that $r_N(n,l) \approx
e^{q_N(n,l)}$, where:
\begin{eqnarray}
q_N(n,l) &=& (N-n)\ln{[(N-n)/e]}-l\ln{(l/e)}-(N-n-l)\ln{[(N-n-l)/e]} \nonumber \\
&+&(N-n-l)\ln{\left(1-p(1-p)^{n-1}\right)}+l[\ln{p}+(n-1)ln{(1-p)}] \label{qu}
\end{eqnarray}
To calculate the largest contributor under the sum in (\ref{exact}) we use the
saddle point method: $\int dx e^{f(x)} \approx \sqrt{2\pi} e^{f(x_0)}/\sqrt{-f''(x_0)}$
where $f'(x_0) = 0$. In our case
we thus need to consider:
\begin{equation}
\left. \frac{\partial q_N(n,l)}{\partial n}\right|_{n^{*}(l)}=0 \label{dn1}
\end{equation}
where $n^{*}(l)$ denotes the index of the maximal contributor for a given $l$.
The difficulty we get into by trying to find $n^*(l)$ from (\ref{dn1})
is that the equation cannot be solved explicitly for $n^*(l)$. To get around this,
let us consider instead the derivative:
\begin{equation}
\left. \frac{\partial q_N(n,l)}{\partial l}\right|_{\hat{l}(n)}=0 \label{dl1}
\end{equation}
defining $\hat{l}(n)$. Performing the derivation the solution is easily found
as:
\begin{equation}
\hat{l}(n)=(N-n)p(1-p)^{n-1}\;. \label{lhat}
\end{equation}
Since $\hat{l}(n)$ is a monotonic function of $n$, it is invertible.
($\hat{l}'(n) < 0$). The inverse of
(\ref{lhat}), will be denoted by $\hat{n}(l)$. This means that:
\begin{equation}
\left. \frac{\partial q_N(n,l)}{\partial l}\right|_{\hat{n}(l)}=0\;. \label{dn2}
\end{equation}
Next, we observe that  $\hat{l}(n)$ satisfies (\ref{dn1}) when inserting it
into its explicit expression. Accordingly, it will also be satisfied by
$\hat{n}(l)$. Assuming that there is only one solution to (\ref{dn1}) it thus
follows that:
\begin{equation}
n^*(l)=\hat{n}(l)\;. \label{neqs}
\end{equation}
If we now calculate $q_N(n,l)$, at the saddle point, we find that
$q_N(n^*(l),l)=0$ (using the fact that the parametric curve of the maximum can
be written as either $(n^*(l),l)$ or $(n,\hat{l}(n))$ and thus calculating
$q_N(n,\hat{l}(n))$). This means that we need to go one step further in the
Stirling series, in order to calculate the leading
piece of $e^{\ln{r_N(n,l)}}$ at the saddle point. For the saddle point itself, we
use the same expression as previously (obtained with the first order Stirling approximation)
because as it can be shown, the corrections introduced by the next term in the
Stirling approximation are vanishing as $N \to \infty$ and therefore they will be neglected. Thus using the
next order term in as well in the Stirling series ($\ln{(x!)}\approx x(\ln{x} - 1)-\ln{(\sqrt{x})}+
\ln{(\sqrt{2\pi})}$) and writing
\begin{equation}
\ln{r_N(n,l)} \approx q_N(n,l)+s_N(n,l)\;, \label{rnqs}
\end{equation}
where $s_N(n,l)$ is the correction generated this way,
we obtain:
\begin{equation}
e^{s_N(n^*(l),l)} = \frac{1}{\sqrt{2\pi}} \sqrt{\frac{N-n^*(l)}{N-n^*(l)-l}}
\frac{1}{\sqrt{l}}=\frac{1}{\sqrt{2\pi}}\frac{1}{\sqrt{l}}+{\cal O}
\left(\frac{\ln{z}}{z}\right)\;. \label{sqs}
\end{equation}
Calculating the second derivative $\partial^2 q_N(n,l)/\partial n^2$ at the point
(\ref{lhat}), one finally obtains:
\begin{equation}
\frac{\partial^2 q_N(n,l)}{\partial n^2}=-l \frac{z^2}{N^2} -
l \mathcal{O}\left(\frac{z}{N^2}\right) -
l \mathcal{O}\left(\frac{z^3}{N^3}\right) \label{sdv}
\end{equation}
Combining (\ref{sdv}) with (\ref{rnqs}), (\ref{sqs}) in the saddle point formula,
one obtains that:
\begin{equation}
R(l)\approx \frac{1}{z l}, \label{arlz}
\end{equation}
valid in the domain $1 \leq l \leq l_{c}$. The cutoff value $l_c$ is
determined by the validity range of the saddle-point method: since the
function $n^*(l)$ is monotonically decreasing, at $l=l_{c}$
it will hit the lowest allowed value by the range of the integral
(or sum), namely, at $n^*(l_{c}) = 1$. Since $\hat{l}$ is the inverse
function of $n^*$, it follows that
\begin{equation}
l_{c}=\hat{l}(n^*(l_{c}))=\hat{l}(1)=p(N-1)=z
\end{equation}
meaning that the cutoff for the $1/l$ scaling law happens at $z$, which is
indeed confirmed by the numerical simulations shown in Fig. \ref{fig:FIG4}.

\section{Scale-free networks: results of a selection mechanism?}

\begin{figure}[htbp]
\protect\vspace*{-0.1cm} \epsfxsize = 4.5 in
\centerline{\epsfbox{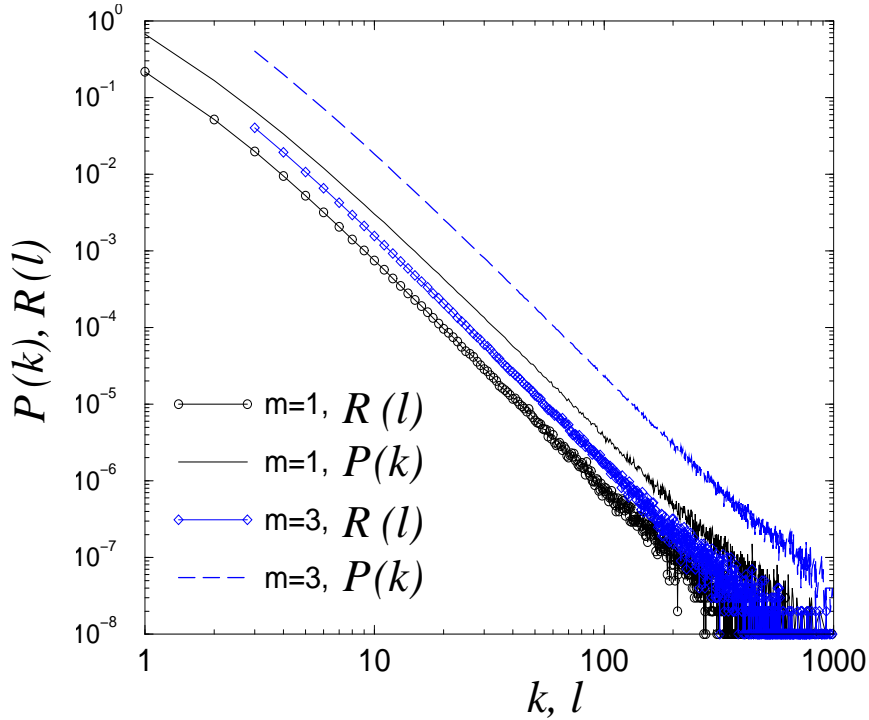}}
\protect\vspace*{-0.2cm}
\caption{
The degree distributions of the gradient network ($R(l)$) and the substrate ($P(k)$), when
the substrate is a BA scale-free graph with parameter $m$ ($m=1,3$).
Here $N=10^5$, and the average was performed over $10^3$ samples.}
\label{fig:FIG5}
\end{figure}

If the substrate graph is a scale-free network
(here we used the Barab\'asi-Albert (BA) process with parameter $m$
to
generate the scale-free network \cite{AB02}, but others will lead to similar
conclusions as long as $\gamma > 2$), the gradient graph will still be a power-law. Here $m$
is the number of ``stubs'' of an incoming node which will attach preferentially
to the existing network in the growth process. Figure \ref{fig:FIG5}
shows the in-degree distribution of the
corresponding gradient networks (for $m=1$ and $m=3$, lines-markers)
which are to be compared with the degree distribution of the substrate network
itself (lines). One immediate conclusion is that the gradient network is the
{\em same type} of structure as the substrate in this case, i.e., it is
a scale-free (power law) graph with the same exponent!

If $N_l^{(in)}$ denotes the number of nodes with $l$ in-links, the
total number of nodes {\em receiving} gradient flow will be
$N_{receive}=\sum_{l \geq 1}N_l^{(in)}$. The total number of
gradient edges generated (total flow) is simply $N_{send} = N$
because every node has exactly one out-link.

\begin{figure}[htbp]
\protect\vspace*{-0.1cm} \epsfxsize = 4 in
\centerline{\epsfbox{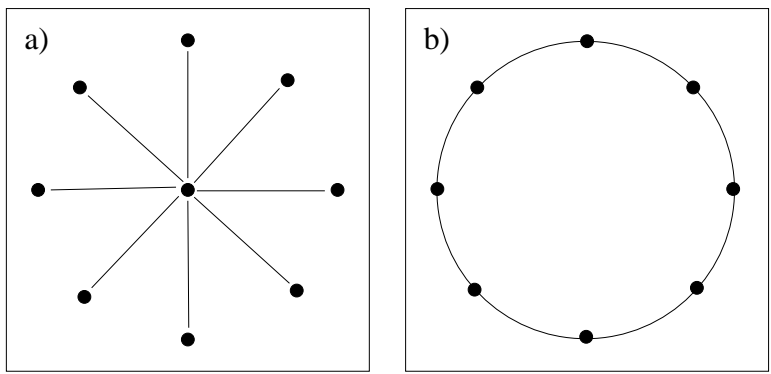}}
\protect\vspace*{-0.2cm}
\caption{Illustrating the influence of structure on flows. In a) the central node
will have maximal congestion, while in b) there is no congestion.}
\label{fig:FIG6}
\end{figure}

If the flow received by a node has to be processed, it will happen
at a finite rate. For example, a node receiving a packet, has to
read off its destination  and find out to which neighbor to send it.
This is a physical process and takes a non-zero amount of time. Thus,
if a node receives too many packets per unit time, they will form
a queue, and long queues will generally cause delays in information
transmission and thus leads to congestion, or jamming.
An important question then arises: Can the topology of the underlying
substrate graph influence the level of congestion in the network?\\
The answer is yes, as illustrated through the following trivial
examples. a) If the network was a star-like structure as in
Fig. \ref{fig:FIG6}a), then obviously all pairs of nodes would be
at most two hops from each other, which is advantageous from the point of
view of shortest distance between sources and destinations (and also
routing would be very simple), however it would not work for large networks,
because the central node would have to handle {\em all} the traffic from the
other nodes and would have to process an extremely large queue.
b) On the contrary, if the network would have a ring-like structure as in
\ref{fig:FIG6}b), then in average there would be one server per one client,
a rather advantageous setup from the point of view of having no congestion,
however, there would be no short distances for transmission. More importantly,
for both structures in Fig. \ref{fig:FIG6}, the network is rather vulnerable:
in case of a) the failure of the central node, and in the case of b) the failure
of any node, would cause a complete breakdown of transmission.

To characterize this interdependence between structure and flow in more generality, we
introduce the ratio $N_{receive}/N_{send}$, which, naturally, will be related to
the instantaneous global {\em congestion} in the network. As explained above, if
this ratio is small, then there will be only a few nodes ($N_{receive}$) processing
the flow of many ($N_{send}$) others  and therefore, long queues are bound to occur, leading to
congestion.

If we assume that flows are generated by gradients in the network, we can define:
\begin{equation}
J = 1-\left\langle \left\langle \frac{N_{receive}}{N_{send}}
\right\rangle_{h} \right\rangle_{nw} = R(0) \label{jam}
\end{equation}
as the {\em congestion (or jamming) factor}. Certainly, $J=1$
means maximal congestion and $J=0$ corresponds to no congestion, and
we always have $J \in [0,1]$.
Note that $J$ is rather a {\em congestion pressure} characteristic
generated by gradients, than an actual throughput characteristic.

For the random graph substrate $G_{N,p}\;$,
$J^{R}(N,p)=\frac{1}{N}\sum_{n=1}^{N-1}\left[1-p(1-p)^n \right]^{N-1-n}$.
We can show that in the limit $p = const.$ and $N \to \infty$,
$J^{R}(N,p)=1-\frac{\ln N}{N \ln \left(\frac{1}{1-p}\right)}\left[
1+{\cal O}\left(\frac{1}{N}\right)\right] \to 1$, i.e., the random graph
becomes {\em maximally congested}. When $z=N p$ is kept constant while
$N\to \infty$, a good approximation is $J^{R}(z) = \lim_{N\to \infty} J^{R}(N,z/N)
=\int_{0}^1 dx\; e^{-z(1-x)e^{-zx}}$. The $J^{R}(z)$ function has a minimum
at $z^* = 2.8763...$, when $J^{R}(z^*) = 0.6295..$ and $J^{R}(0)=1$. Since
$e^{zxe^{-zx}}$ is always bounded:
$e^{zxe^{-zx}}\in\left[1, e^{1/e}\right]$, for all
$x \geq 0$ and $z \geq 0$, $J^{R}(z) \geq \int_{0}^1 dx\; e^{-z e^{-zx}}
=\frac{1}{z}\left[Ei(-z)-Ei\left( -z e^{-z}\right) \right]$. Expanding for
$z >> 1$, we obtain:
$J^{R}(z) \geq 1 - \frac{\ln z+ \mbox{\bf \mbox{ \em \tiny C}}}{z}-... \to 1$,
($\mbox{\bf \mbox{\em C}}$ is the Euler constant) i.e.,
the random graph asymptotically becomes {\em maximally congested, or jammed}.
The latter result (up to corrections in $1/z$) can immediately be obtained if one
uses directly the asymptotic form (\ref{arlz}) and the fact that $R(0)\approx 1-
\int_{1}^{z} dl R(l)$.
\begin{figure}[htbp]
\protect\vspace*{-0.0cm} \epsfxsize = 5 in
\centerline{\epsfbox{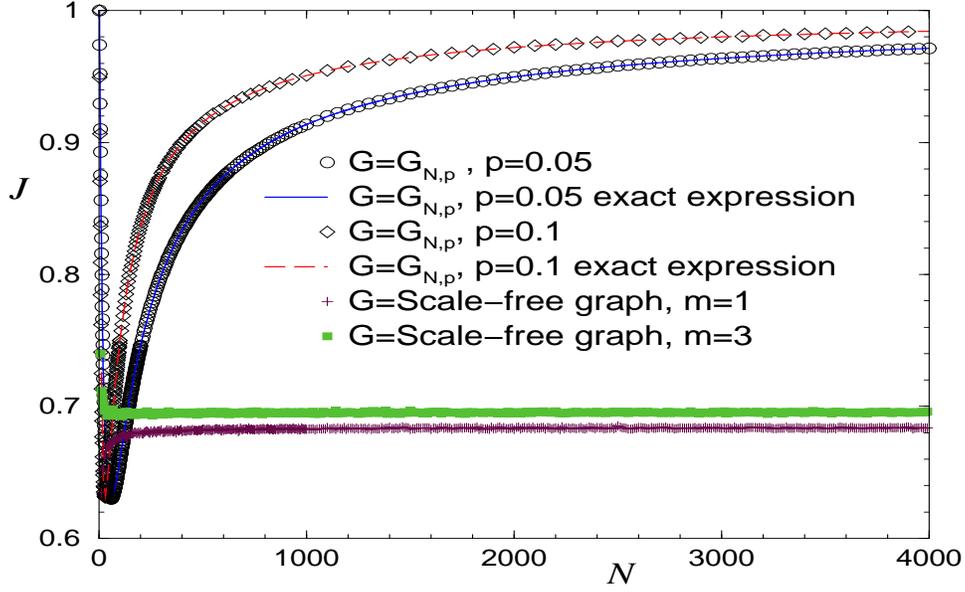}}
\protect\vspace*{-0.2cm}
\caption{The congestion coefficient for random graphs
($\circ$, $p=0.05$; $\diamond$, $p=0.1$) and
scale-free networks ($+$, $m=1$; $\square$, $m=3$).}
\label{fig:gnFig4}
\end{figure}

For scale-free networks, however, the conclusion
about jamming is drastically different.
We find that  the jamming (pressure)
coefficient $J$ is always a {\em constant,
independent of $N$}, in other words,
scale-free networks {\em are not prone to maximal
jamming}! In particular, $J^{SF}(N,m=1) = 0.6839...$ and
$J^{SF}(N,m=3) = 0.6955...$
Figure \ref{fig:gnFig4} shows as comparison
the congestion factors as function
of network size both for random graph and
scale-free network substrates.

Most real-world networks evolve more-or-less spontaneously
(like the internet or www) and they also can reach massive proportions
(order $10^8$ nodes). At such proportions, pure random graph structures
would generate maximal congestion pressure (practically equal to unity)
in the network and thus such substrates would be very inefficient for
transport. Scale-free networks, however, have a congestion pressure
which is a constant bounded well away from unity, and thus they are rather
efficient substrates for transport. So why not all real-world networks are
scale-free? Our analysis assumed that all edges have the same transport
properties (conductance, or "cost"), which is true for some networks
like the www, the internet, and not true for others: power grid, social networks,
etc. When there are weights on the {\em edges} (conductance or cost) the
actual transport efficiency (determined by actual throughput) will strongly
depend on those and therefore also select the network topology.
Note that the congestion pressure depends on {\em local properties}
of network topology (2-step neighborhoods). Thus we expect that all
networks with similar 2-step neighborhood distributions
would have similar gradient networks, not just the models studied here.

In summary, after introducing the concept of gradient flow networks,
we have shown why certain complex networks might emerge to be scale-free.
We do not give a specific mechanism
for network evolution, which we believe to be network and process dependent,
and therefore not universal.
Instead, the network evolution mechanism is {\em selected}
such, that the transport (the network's main function) is
efficient, while  a number of constraints imposed by the
specific nature of the network (edge cost, conductance, etc.) are
obeyed.

In the case of the internet, if a router is constantly jammed,
engineers will resolve the problem by connecting to the network other routers
in its network vicinity, to share the load. This is a tendency to
{\em optimize} the network flow {\em locally}. A series of such local optimization
processes will necessarily have to constrain the global structure of the network.
It seems that scale-free networks are within the class of networks obeying
this type of constraint.

\subsection*{Acknowledgements}

The authors acknowledge useful discussions with M. Anghel,
B. Bollob\'as, P.L. Erd\H{o}s, E.E. Lopez and D. Sherrington.
Z.T. is supported by the US DOE
under contract W-7405-ENG-36, K.E.B. is supported by the
NSF through DMR-0406323, DMR-0427538, and the Alfred P. Sloan
Foundation. B. K. and G. K. were supported in part by NSF
through DMR-0113049, DMR-0426488, and the Research Corp. Grant No. RI0761.
B.K. was also supported in part by the LANL summer student program in
2004 through US DOE Grant No. W-7405-ENG-36.

\appendix

\section{An analytic derivation of the in-degree distribution}

When calculating the degree distribution,
we have to perform two averages:
one corresponding to the scalar field disorder
\begin{equation}
\langle\bullet\rangle_h = \int dh_0 ... dh_{N-1}\;
\eta(h_0) .. \eta(h_{N-1}) \bullet\;, \label{fieldav}
\end{equation} and the other to an
average over the network (graph ensemble):
\begin{equation}
\langle\bullet\rangle_{nw} = \sum_{a_{01}}...\!\!\sum_{a_{N-2 N-1}}
v(a_{01})...v(a_{N-2 N-1}) \bullet\;, \label{nta}
\end{equation}
where $v(a) = p^{a}(1-p)^{1-a}$,
$a\in\{0,1\}$ and $\sum_a \equiv \sum_{a=0}^{1}$.
Here $G$ is the binomial random graph $G_{N,p}$ with
$N$ nodes and link-probability $p$.
The integrals in (\ref{fieldav}) are computed over the range
of the scalar field and the summation in (\ref{nta}) is over all
$N(N-1)/2$ pairs $(i,j)$ with $i < j\;$.

\begin{figure}[htbp]
\protect\vspace*{-0.1cm} \epsfxsize = 2.4 in
\centerline{\epsfbox{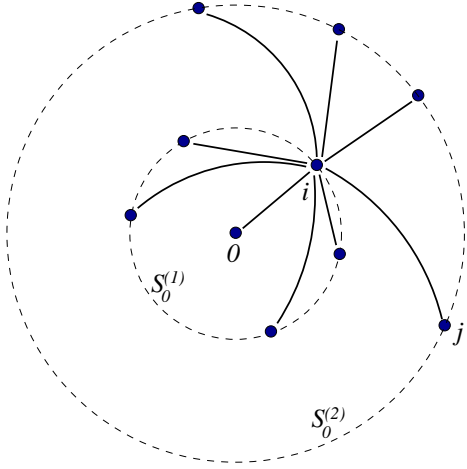}}
\protect\vspace*{-0.2cm}
\caption{Node 0 has a gradient edge from $i$, if its scalar value is larger
than the scalars of all its neighbors $j\neq 0$.} \label{fig:FIG3}
\end{figure}

In order to calculate the in-degree distribution, we
define first a counter operator for the in-links.
Without restricting the generality we calculate the in-degree
of the gradient network for node 0 namely, $k_{0}^{(in)}$.
Let us introduce ${\bf B} = {\bf I}+{\bf A}$,
where ${\bf I}$ is the $N\times N$ identity matrix
so $b_{ij} = \delta_{i,j}+a_{ij}$, and the quantities:
$H_i(j) = 1 - b_{ij} + b_{ij}\;\theta(h_0 - h_j)$
for $i,j \in V$, and $i \in S^{(1)}_0$.  Thus, the in-link counter
can be written as:
\begin{equation}
k_0^{(in)} = \sum_{i=1}^{N-1}a_{0i}\prod_{j=1}^{N-1}
H_i(j) \;.\label{k0in}
\end{equation}
With the aid of Fig. \ref{fig:FIG3} we see that indeed this expression
will count the number of gradient edges into node 0: $H_i(j)$
is zero only if the {\em  neighbor} $j$ of $i$ (except node $0$)
has a larger scalar value than node 0, i.e., $h_0 < h_j$, otherwise
$H_i(j)$ is equal to unity. Therefore a term under the sum in (\ref{k0in})
will be non-zero if and only if for {\em all} neighbors $j$ of $i$
(i.e., $b_{ij}=1$)  $h_j < h_0$ holds, making the edge $(i,0)$ to be the gradient
edge for node $i$.

The probability that a node will have $l$ in-degree on
the gradient network ${\bf F}$, is:
\begin{equation}
R(l)=
\left\langle \left\langle
\delta_{l, \;k^{(in)}_0}
\right\rangle_{h} \right\rangle_{nw} =
\int\limits_{-\pi}^{\pi}\frac{dq}{2 \pi} e^{{\mathbf i}ql}
\left\langle \left\langle
e^{-{\mathbf i}q k^{(in)}_0}
\right\rangle_{h} \right\rangle_{nw} \;. \label{degd}
\end{equation}
First, we compute the average over the scalar field. (The
order of the averages does not matter, however, it is formally easier this way.)
Let us denote
\begin{equation}
L_G(q) = \left\langle e^{-{\mathbf i}qk^{(in)}_0} \right\rangle_h \label{LGq}
\end{equation}
We have:
\begin{equation}
L_G(q) = \int dh_0...\int dh_{N-1}\;
\eta(h_0) .. \eta(h_{N-1})
e^{-{\mathbf i}q \sum_{i=1}^{N-1}a_{0i}\prod_{j=1}^{N-1}
\left[ 1 - b_{ij} + b_{ij}\;\theta(h_0 - h_j) \right]}
\end{equation}
Let
$M_i(m)=\prod_{j=1}^{m} H_i(j)$.
So
\begin{equation}
L_G(q) = \int dh_0...\int dh_{N-1}\;
\eta(h_0) .. \eta(h_{N-1})
e^{-{\mathbf i}q \sum_{i=1}^{N-1}a_{0i}M_i(N-1)}\;.
\end{equation}
Using the recursion
\begin{equation}
M_i(m) = \left[ 1-b_{im} + b_{im}\;\theta(h_0 - h_m) \right] M_i(m-1)\;,
\end{equation}
the integral over $h_{N-1}$ can be performed:
\begin{eqnarray}
L_G(q)=&&\int\!\! dh_0...\int dh_{N-2}\;
\eta(h_0) .. \eta(h_{N-2}) \nonumber \\
&&\times \left\{ \gamma(h_0) e^{-{\mathbf i}q \sum_{i=1}^{N-1}a_{0i}M_i(N-2)} +
[1-\gamma(h_0)]e^{-{\mathbf i}q \sum_{i=1}^{N-1}a_{0i}[1-b_{i N-1}]M_i(N-2)}\right\}
\end{eqnarray}
where $\gamma(x)=\int^{x}dy \;\eta(y)$.
Performing all the integrals recursively, except for $h_0$, we obtain:
\begin{equation}
L_G(q)=\sum_{n=0}^{N-1} J(N,n)\!\!\!\!
\sum_{[\tau]_n \in {\cal P}_n(N-1)}\!\!\!\!
e^{-{\mathbf i}q\sum_{i=1}^{N-1}a_{0i}\prod_{j=1}^n (1-b_{i\tau(j)})} \nonumber
\end{equation}
where $J(N,n) = \int dh_0 \eta(h_0)
[\gamma(h_0)]^{N-1-n} [1-\gamma(h_0)]^{n}$.
Here $[\tau]_n = \{ \tau(1), \tau(2), ..., \tau(n) \}$ is a $n$-subset
of the set $\{ 1,2,...,N-1 \}$ and
${\cal P}_n(N-1)$ denotes the set of all $n$-subsets of
$\{ 1,2,...,N-1 \}$.
We have $|{\cal P}_n(N-1)| = {N-1 \choose n}$.
After a change of variables $u = \gamma(h_0)$ and using
$du = d\gamma(h_0) = \eta(h_0) h_0$ the integral $J(N,n)$ yields
$J(N,n) = \frac{1}{N}{N-1 \choose n }^{-1}$, i.e., the in-degree
distribution is independent on the choice of the $\eta(h)$ distribution!

In the following, we perform the network average
$\left\langle L_G(q) \right\rangle_{nw}$.
For a fixed $n$-subset
$[\tau]_n$, let us denote:
\begin{equation}
Z_{[\tau]_n}(q) \equiv \left\langle
e^{-{\mathbf i}q\sum_{i=1}^{N-1}a_{0i}
\prod_{j=1}^n (1-b_{i\tau(j)})} \right\rangle_{nw}\;. \label{ztq}
\end{equation}
Thus,
\begin{equation}
\left\langle L_G(q) \right\rangle_{nw} =
\frac{1}{N} \sum_{n=0}^{N-1} {N-1 \choose n }^{-1}
\sum_{[\tau]_n \in {\cal P}_n(N-1)} Z_{[\tau]_n}(q)\;. \label{lgq}
\end{equation}
Let
\begin{equation}
T_n = [\tau]_n\cup\bigcup_{j=1}^{n}S^{(1)}_{\tau(j)}
\end{equation}
be the set of vertices $[\tau]_n$ and its neighbors on $G$.

Note, that $\prod_{j=1}^n (1-b_{i\tau(j)}) = 1$ if and only if
$i \not\in T_n$ otherwise it is zero. Therefore,
\begin{equation}
\sum_{i=1}^{N-1}a_{0i}
\prod_{j=1}^n (1-b_{i\tau(j)}) = \mbox{the nr. of neighbors
of 0 {\em which do not belong to} }\;\;T_n.
\end{equation}
From (\ref{nta})
\begin{equation}
 Z_{[\tau]_n}(q) = \sum_{a_{01}}...\!\!\sum_{a_{N-2 N-1}}
v(a_{01})...v(a_{N-2 N-1})\;\prod_{i=1}^{N-1}
e^{-{\mathbf i}q a_{0i}
\prod_{j=1}^n (1-b_{i\tau(j)})}
\end{equation}
Since $\tau(j) \neq 0$, ($[\tau]_n \in {\cal P}_n(N-1)$), the sums
over the matrix variables $a_{0i}$ can be performed:
\begin{equation}
\sum_{a_{0i}} v(a_{0i}) e^{-{\mathbf i}q a_{0i}
\prod_{j=1}^n (1-b_{i\tau(j)})} = 1-p + p
e^{-{\mathbf i}q\prod_{j=1}^n (1-b_{i\tau(j)})}\;,
\end{equation}
and therefore
\begin{equation}
Z_{[\tau]_n}(q) = \sum_{a_{12}}...\!\!\sum_{a_{N-2 N-1}}
v(a_{12})...v(a_{N-2 N-1})\;\prod_{i=1}^{N-1}
\left[ 1-p + p
e^{-{\mathbf i}q\prod_{j=1}^n (1-b_{i\tau(j)})} \right]\;. \label{ztqe}
\end{equation}

The set of vertices
$\{1,2,...,N-1\}$ is split into two groups: $[\tau]_n$ and its
complementary in $\{1,2,...,N-1\}$. Without changing anything,
we can rename the vertices, such that $\{1,2,...,n\} = [\tau]_n$
and $C_{[\tau]_n}=\{n+1,n+2,...,N-1\}$ be the complementary set of $[\tau]_n$.
It is easy to see that only cross-terms ($a_{ij}$ involving one node $i$ from
$[\tau]_n$ and one node $j$ from $C_{[\tau]_n}$)
give non-trivial contribution (i.e., different from unity) in (\ref{ztqe}). Thus:
\begin{equation}
Z_{[\tau]_n}(q) =
\prod_{i=n+1}^{N-1}\sum_{a_{1 i}}...\sum_{a_{n i}}
v(a_{1 i})...v(a_{n i})
\left[ 1-p + p
e^{-{\mathbf i}q\prod_{j=1}^n (1-a_{ji})} \right] \label{zst}
\end{equation}
Let $\alpha_1 = 1-p$ and $\beta_1 = p$. Then:
\begin{equation}
\sum_{a_{1 i}}v(a_{1 i})
\left[\alpha_1+\beta_1 e^{-{\mathbf i}q(1-a_{1i})...(1-a_{ni})}\right] =
\alpha_2+\beta_2 e^{-{\mathbf i}q(1-a_{2i})...(1-a_{ni})}
\end{equation}
where $\alpha_2=(1-p)\alpha_1+p$ and $\beta_2 = (1-p)\beta_1$. The summation
over the rest of the matrix elements can be similarly performed to give
(for a fixed node $i$)
\begin{equation}
\alpha_{n+1}+\beta_{n+1}e^{-{\mathbf i}q}\;. \label{abn}
\end{equation}
The coefficients are determined from the recursion:
\begin{eqnarray}
\left\{
\begin{array}{ll}
\alpha_k = (1-p)\alpha_{k-1}+p\;, & \;\;\alpha_1 = 1-p \\
\beta_k = (1-p)\beta_{k-1}\;, & \;\;\beta_1 = p
\end{array} \right.
\end{eqnarray}
which obeys $\alpha_k + \beta_k = 1$ for all $k$.
These recursions are easily solved:
\begin{equation}
\alpha_{n+1} = 1-p(1-p)^n\;,\;\;\;\beta_{n+1} = p(1-p)^n\;.
\end{equation}
Thus (\ref{abn}) becomes
$1-p(1-p)^n+p(1-p)^ne^{-{\mathbf i}q}$.
Since for all indices $i$ in (\ref{zst}) the result of the summations
is the same, one finally obtains:
\begin{equation}
Z_{[\tau]_n}(q) =\left[ 1-p(1-p)^n\left(1-e^{-{\mathbf i}q}\right)
\right]^{N-1-n}\;. \label{zf}
\end{equation}
Because the result in (\ref{zf}) is not specific of the $[\tau]_n$
set, for all realizations of $[\tau]_n$, $Z_{[\tau]_n}(q)$ is
the same expression, and thus the sum over all realizations of
$[\tau]_n$ in (\ref{lgq}) will generate the factor
$|{\cal P}_n(N-1)| = {N-1 \choose n}$
which cancels the combinatorial factor in (\ref{lgq}). Thus:
$\left\langle L_G(q) \right\rangle_{nw}=
\frac{1}{N}\sum_{n=0}^{N-1}Z_{[\tau]_n}(q)$. Plugging this into (\ref{degd}), and
performing the integral over the $q$ variable we obtain:
\begin{equation}
R(l) = \frac{1}{N}\sum_{n=0}^{N-1}\;
{ N-1-n \choose l} \left[1-p(1-p)^n\right]^{N-1-n-l}
\left[ p(1-p)^n\right]^l  \label{RNl3}
\end{equation}
with the usual convention ${ M \choose m} = 0$ for $M < m$.
Equation (\ref{RNl3}) is the exact expression for the in-degree
distribution of the gradient network $\nabla G$.

\end{document}